\documentclass[journal]{IEEEtran}
\ifCLASSINFOpdf
\else
\fi
\usepackage{epsfig}
\usepackage{graphicx}
\usepackage{amsmath}
\usepackage{amssymb}
\usepackage{verbatim}
\usepackage{mathrsfs}
\usepackage{cite}
\usepackage[T1]{fontenc}

\begin{document}
%
\title{Power of Deep Learning for Channel Estimation and Signal Detection in OFDM Systems }

\author{\IEEEauthorblockN{Hao Ye, Geoffrey Ye Li, \emph{Fellow, IEEE}, and Biing-Hwang Fred Juang, \emph{Fellow, IEEE}\\}
\thanks{The authors are with the Department
of Electrical and Computer Engineering, Georgia Institute of Technology, Atlanta,
GA, 30332 USA e-mail: (yehao@gatech.edu; liye@ece.gatech.edu; juang@ece.gatech.edu).}
}

\maketitle
\begin{abstract}

This article presents our initial results in deep learning for channel estimation and signal detection in orthogonal frequency-division multiplexing (OFDM) systems.
In this article, we exploit deep learning to handle wireless OFDM channels in an end-to-end manner.
Different from existing OFDM receivers that first estimate channel state information (CSI) explicitly and then detect/recover the transmitted symbols using the estimated CSI, the proposed deep learning based approach estimates CSI implicitly and recovers the transmitted symbols directly.
To address channel distortion, a deep learning model is first trained offline using the data generated from simulation based on channel statistics and then used for recovering the online transmitted data directly.
From our simulation results, the deep learning based approach can address channel distortion and detect the transmitted symbols with performance comparable to the minimum mean-square error (MMSE) estimator.
Furthermore, the deep learning based approach is more robust than conventional methods when fewer training pilots are used, the cyclic prefix (CP) is omitted, and nonlinear clipping noise exists.
In summary, deep learning is a promising tool for channel estimation and signal detection in wireless communications with complicated channel distortion and interference.
\end{abstract}
\IEEEpeerreviewmaketitle
\section{Introduction}

Orthogonal frequency-division multiplexing (OFDM) is a popular modulation scheme that has been widely adopted in wireless broadband systems to combat frequency-selective fading in wireless channels.
Channel state information (CSI) is vital to coherent detection and decoding in OFDM systems. Usually, the CSI can be estimated by means of pilots prior to the detection of the transmitted data.  With the estimated CSI, transmitted symbols can be recovered at the receiver.

Historically, channel estimation in OFDM systems has been thoroughly studied. The traditional estimation methods, i.e., least square (LS) and minimum mean-square error (MMSE), have been utilized and optimized in various conditions \cite{Li}.
The method of LS estimation requires no prior channel statistics, but its performance may be inadequate.
The MMSE estimation in general leads to much better detection performance by utilizing the second order statistics of channels.

In this article, we introduce a deep learning approach to channel estimation and symbol detection in an OFDM system. Deep learning and artificial neural networks (ANNs) have numerous applications. In particular, it has been successfully applied in localization based on CSI \cite{deeploc}, channel equalization \cite{mlpeq}, and channel decoding \cite{ChannelDecoding} in communication systems.
With the improving computational resources on devices and the availability of data in large quantity, we expect deep learning to find more applications in communication systems.

ANNs have been demonstrated for channel equalization with online training, which is to adjust the parameters according to the online pilot data.
However, such methods can not be applied directly since, with deep neural networks (DNNs), the number of parameters becomes much increased, which requires a large number of training data together with the burden of a long training period.
To address the issue, we train a DNN model that predicts the transmitted data in diverse channel conditions. Then the model is used in online deployment to recover the transmitted data.

This article presents our initial results in deep learning for channel estimation and symbol detection in an end-to-end manner. It demonstrates that DNNs have the ability to learn and analyze the characteristics of wireless channels that may suffer from nonlinear distortion and interference in addition to frequency selectivity.
To the best of our knowledge, this is the first attempt to use learning methods to deal with wireless channels without online training.
The simulation results show that deep learning models achieve performance comparable to traditional methods if there are enough pilots in OFDM systems, and it can work better with limited pilots, channel interference, and nonlinear noise.
Our initial research results indicate that deep learning can be potentially applied in many directions in signal processing and communications.


\section{Deep Learning Based Estimation and Detection}
In this section, we present a method where deep learning is exploited as an end-to-end approach for channel estimation and symbol detection.
The DNN model is trained based on simulated data offline, which views OFDM and the wireless channel as complete black boxes.

\subsection{Deep Learning Methods}
\begin{figure}[!t]
\centering
\includegraphics[width = 0.9\linewidth]{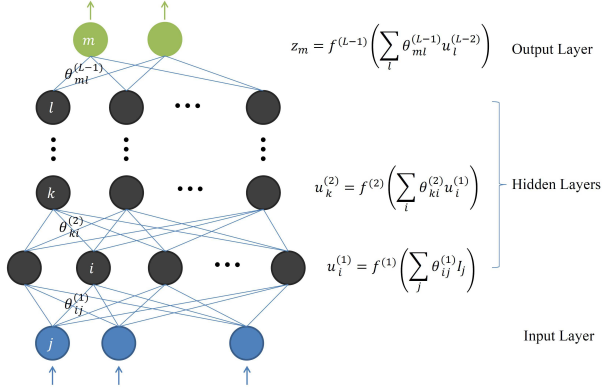}
\caption{An example of deep learning models. }\label{fig:deep}
\end{figure}
Deep learning has been successfully applied in a wide range of areas with significant performance improvement, including computer vision \cite{Imagenet}, natural language processing \cite{DL_NLP}, speech recognition \cite{DL_Speech}, and so on.
A comprehensive introduction to deep learning and machine learning can be found in \cite{DLIntro}.

The structure of a DNN model is shown in Fig. \ref{fig:deep}.
Generally speaking, DNNs are deeper versions of ANNs by increasing the number of hidden layers in order to improve the ability in representation or recognition.
Each layer of the network consists of multiple neurons, each of which has an output that is a nonlinear function of a weighted sum of neurons of its preceding layer, as shown in Fig. \ref{fig:deep}.
The nonlinear function may be the Sigmoid function, or the Relu function, defined as
$f_{S}(a) = \frac{1}{1+e^{-a}}$, and $f_{R}(a) = \max(0,a)$, respectively.
Hence, the output of the network $\mathbf{z}$ is a cascade of nonlinear transformation of input data $\mathbf{I}$, mathematically expressed as
\begin{equation}
\mathbf{z} = f(\mathbf{I}, \boldsymbol{\theta}) = f^{(L-1)}(f^{(L-2)}(\cdot \cdot \cdot f^{(1)}(\mathbf{I}))),
\end{equation}
where $L$ stands for the number of layers, and $\boldsymbol{\theta}$ denotes the weights of the neural network.
The parameters of the model are the weights for the neurons, which need to be optimized before the online deployment.
The optimal weights are usually learned on a training set, with known desired outputs.

\subsection{System Architecture}
\begin{figure}[!th]
\centering
\includegraphics[width=0.9\linewidth]{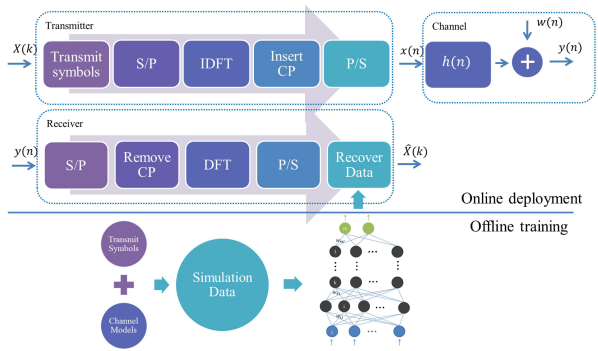}
\caption{System model. }\label{fig:systemmodel}
\end{figure}
The architecture of the OFDM system with deep learning based channel estimation and signal detection is illustrated in Fig. \ref{fig:systemmodel}.
The baseband OFDM system is the same as the conventional ones.
On the transmitter side, the transmitted symbols inserted with pilots are first converted to a paralleled data stream, then the inverse discrete Fourier transform (IDFT) is used to convert the signal from the frequency domain to the time domain.
After that, a cyclic prefix (CP) is inserted to mitigate the inter-symbol interference (ISI). The length of the CP should be no shorter than the max delay spread of the channel.

We consider a sample-spaced multi-path channel described by complex random variables ${\{h(n)\}}_{n=0}^{N-1}$. Thus the received signal, $y(n)$, can be expressed as
\begin{equation}
y(n) = x(n)\otimes h(n) + w(n),
\end{equation}
where $\otimes$ denotes the circular convolution while $x(n)$ and $w(n)$ represent the transmitted signal and the additive white Gaussian noise (AWGN), respectively.
After removing the CP and performing DFT, the received frequency domain signal is
\begin{equation}
Y(k) = X(k)H(k) + W(k),
\end{equation}
where $Y(k)$, $X(k)$, $H(k)$, and $W(k)$ are the DFT of $y(n)$, $x(n)$, $h(n)$ and $w(n)$, respectively.

We assume that the pilot symbols are in the first OFDM block while the following OFDM blocks consist of the transmitted data. Together they form a \emph{frame}.
The channel can be treated as constant spanning over the pilot block and the data blocks, but change from one frame to another.
The DNN model takes as input the received data consisting of one pilot block and one data block in our initial study, and recovers the transmitted data in an end-to-end manner.

As shown in Fig. \ref{fig:systemmodel}, to obtain an effective DNN model for joint channel estimation and symbol detection, two stages are included.
In the offline training stage, the model is trained with the received OFDM samples that are generated with variant information sequences and under diverse channel conditions with certain statistical properties, such as typical urban or hilly terrain delay profile.
In the online deployment stage, the DNN model generates the output that recovers the transmitted data without explicitly estimating the wireless channel.

\subsection{Model Training}
The models are trained by viewing OFDM modulation and the wireless channels as black boxes.
Historically, researchers have developed many channel models for CSI that well describe the real channels in terms of channel statistics.
With these channel models, the training data can be obtained by simulation.
In each simulation, a random data sequence is first generated as the transmitted
symbols and the correspondent OFDM frame is formed with pilot symbols.
The current random channel state is simulated based on the channel models. The received OFDM signal is obtained based on the OFDM frames undergoing the current channel distortion, including the channel noise.
The received signal and the original transmitted data are collected as the training data.
The model is trained to minimize the difference between the output of the neural network and the transmitted data.
The difference can be portrayed in several ways.
In our experiment settings, we choose the $L_2$ loss,
\begin{equation}
L_2 = \frac{1}{N}\sum_k(\hat{X}(k)-X(k))^2,
\end{equation}
where $\hat{X}(k)$ is the prediction and $X(k)$ is the supervision message, which is the transmitted symbols in this situation.


The DNN model we use consists of five layers, three of which are hidden layers.
The numbers of neurons in each layers are $256$, $500$, $250$, $120$, $16$.
The input number corresponds to the number of real parts and imaginary parts of $2$ OFDM blocks that contain the pilots and transmitted symbols, respectively.
Every $16$ bits of the transmitted data are grouped and predicted based on a single model trained independently, which is then concatenated for the final output.
The Relu function is used as the activation function in most layers except in the last layer where the Sigmoid function is applied to map the output to the interval $[0,1]$.

\section{Simulation Results}
We have conducted several experiments to demonstrate the performance of the deep learning methods for joint channel estimation and symbol detection in OFDM wireless communication systems.
A DNN model is trained based on simulation data, and is compared with the traditional methods in term of bit-error rates (BERs) under different signal-to-noise ratios (SNRs).
In the following experiments, the deep learning based approach is proved to be more robust than LS and MMSE under scenarios where fewer training pilots are used, the CP is omitted, or there is nonlinear clipping noise.
In our experiments, an OFDM system with $64$ sub-carriers and the CP of length $16$ is considered. The wireless channel follows the wireless world initiative for new radio model (WINNER \uppercase\expandafter{\romannumeral2}) \cite{winner}, where the carrier frequency is $2.6$ GHz, the number of paths is $24$,  and typical urban channels with max delay $16$ are used. QPSK is used as the modulation method.







\begin{figure}[!t]
\centering
\includegraphics[width=0.8\linewidth]{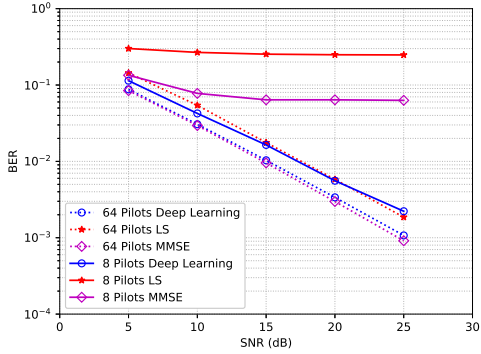}
\caption{BER curves of deep learning based approach and traditional methods.}\label{fig:Compare}
\end{figure}


\subsection{Impact of Pilot Numbers}

The proposed method is first compared with the LS and MMSE methods for channel estimation and detection, when $64$ pilots are used for channel estimation in each frame.
From Fig. \ref{fig:Compare}, the LS method has the worst performance since no prior statistics of the channel has been utilized in the detection.
On the contrary, the MMSE method has the best performance because the second-order statistics of the channels are assumed to be known and used for symbol detection.
The deep learning based approach has much better performance than the LS method and is comparable to the MMSE method.

Since the channel model has a max delay of $16$, it can be estimated with much fewer pilots, leading to better spectrum utilization.
From Fig. \ref{fig:Compare}, when only $8$ pilots are used, the BER curves of the LS and MMSE methods saturate when SNR is above $10$ dB while the deep learning based method still has the ability to reduce its BER with increasing SNR, which demonstrates that the DNN is robust to the number of pilots used for channel estimation.
The reason for the superior performance of the DNN is that the CSI is not uniformly distributed. The characteristics of the wireless channels can be learned based on the training data generated from the model.



\subsection{Impact of CP}
As indicated before, the CP is necessary to convert the linear convolution of the physical channel into circular convolution and mitigate ISI.
But it costs time and energy for transmission. In this experiment, we will investigate the performance with CP remover.

\begin{figure}[!t]
\centering
\includegraphics[width=0.8\linewidth]{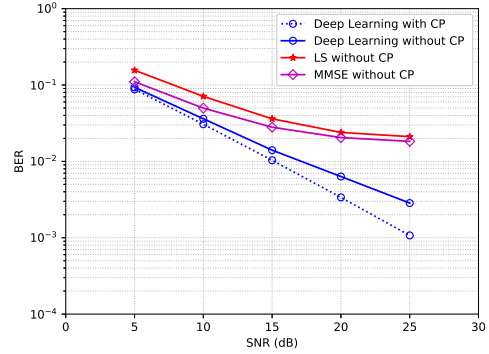}
\caption{BER curves without CP.}\label{fig:CP}
\end{figure}

Fig. \ref{fig:CP} illustrates the BER curves for an OFDM system without CP.
From the figure, neither MMSE nor LS can effectively estimate channel.
The accuracy tends to be saturated when SNR is over $15$ dB.
However, the deep learning method still works well.
This result shows again that the characteristics of the wireless channel have been revealed and can be learned in the training stage by the DNNs.

\subsection{Impact of Clipping and Filtering Distortion}
As indicated in \cite{clipping}, a notable drawback of OFDM is the high peak-to-average power ratio (PAPR).
To reduce PAPR, the clipping and filtering approach serves as a simple and effective approach \cite{clipping}.
However, after clipping, nonlinear noise is introduced that could degrade the estimation and detection performance.
The clipped signal becomes
\begin{equation}
\hat{x}(n) = \begin{cases}
x(n),& \ \text{if } |x(n)| \leq A, \\
A e^{j \phi(n)},& \ \text{otherwise},
\end{cases}，
\end{equation}
where $A$ is the threshold and $\phi(n)$ is the phase of $x(n)$.

\begin{figure}[!t]
\centering
\includegraphics[width=0.8\linewidth]{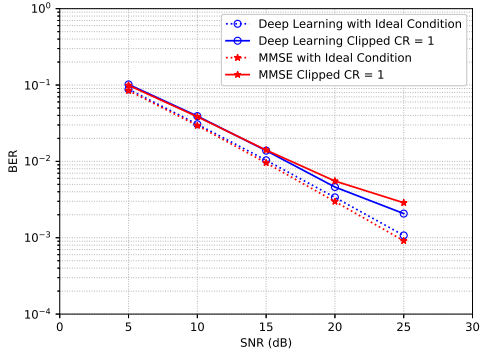}
\caption{BER curves with clipping noise}\label{fig:clipping}
\end{figure}


Fig. \ref{fig:clipping} depicts the detection performance of the MMSE method and deep learning method when the OFDM system is contaminated with clipping noise. From the figure, when clipping ratio (CR = $A/\sigma$, where $\sigma$ is the rms of OFDM signal) is $1$, the deep learning method is better than the MMSE method when SNR is over $15$ dB, proving that deep learning method is more robust to the nonlinear clipping noise.

\begin{figure}[!t]
\centering
\includegraphics[width=0.8\linewidth]{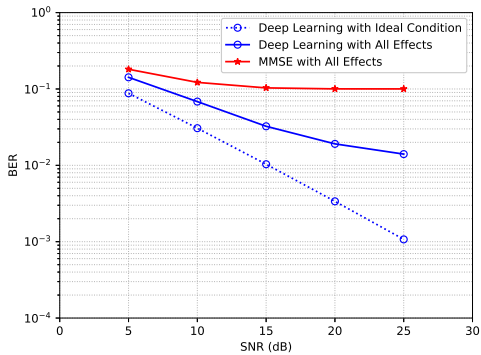}
\caption{BER curves when combining all adversities. }\label{fig:combined}
\end{figure}

Fig \ref{fig:combined} compares DNN with the MMSE method when all above adversities are combined together, i.e., only 8 pilots are used, the CP is omitted, and there is clipping noise.
From the figure, DNN is much better than the MMSE method but has a gap with detection performance under ideal circumstance, as we have seen before.

\subsection{Robustness Analysis}
\begin{figure}[!t]
\centering
\includegraphics[width=0.8\linewidth]{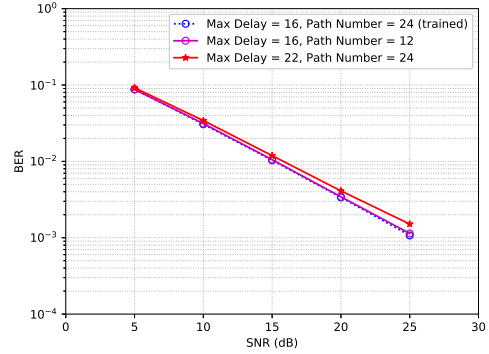}
\caption{\small BER curves with mismatch over training stage and deployment stage. }\label{fig:Robust}
\end{figure}

In the simulation above, the channels in the online deployment stage are generated with the same statistics that are used in the offline training stage.
However, in real-world applications, mismatches may occur between the two stages.
Therefore, it is essential for the trained models to be relatively robust to these mismatches.
In this simulation, the impact of variation in statistics of channel models used during training and deployment stages is analyzed.
Fig \ref{fig:Robust} shows the BER curves when the max delay and the number of paths in the test stage vary from the parameters used in the training stage described in the beginning of this section. From the figure, variations on statistics of channel models do no have significant damage on the performance of symbol detection.

\section{Conclusions}

In this article, we have demonstrated our initial efforts to employ DNNs for channel estimation and symbol detection in an OFDM system.
The model is trained offline based on the simulated data that view OFDM and the wireless channels as black boxes.
The simulation results show that the deep learning method has advantages when wireless channels are complicated by serious distortion and interference, which proves that DNNs have the ability to remember and analyze the complicated characteristics of the wireless channels.
For real-world applications, it is important for the DNN model to have a good generalization ability so that it can still work effectively when the conditions of online deployment do not exactly agree with the channel models used in the training stage.
An initial experiment has been conducted in this article to illustrate the generalization ability of DNN model with respect to some parameters of the channel model.
More rigorous analysis and more comprehensive experiments are left for the future work.
In addition, for practical use, samples generated from the real wireless channels could be collected to retrain or fine-tune the model for better performance.



\begin{thebibliography}{1}


\bibitem{DLIntro}
J.~Schmidhuber, ``Deep learning in neural networks: An overview,'' \emph{Neural
  Networks}, vol. 61, pp. 85--117, Jan. 2015.

\bibitem{Li}
Y.~G.~Li, L.~J.~Cimini, and N.~R.~Sollenberger, ``Robust channel estimation
for OFDM systems with rapid dispersive fading channels,'' \emph{ IEEE
Trans. Commun.}, vol. 46, no. 7, pp. 902–-915, Jul. 1998.



\bibitem{deeploc}
X.~Wang, L.~Gao, S.~Mao and S.~Pandey, 2017. ``CSI-based fingerprinting for indoor localization: A deep learning approach,'' \emph{IEEE Trans. Veh. Technol.}, vol. 66, no. 1, pp. 763--776, Jan. 2017.

\bibitem{ChannelDecoding}
E.~Nachmani, Y.~Beery, and D.~Burshtein, ``Learning to decode linear codes using deep learning,'' \emph{54'th Annual Allerton Conf. On Commun., Control and Computing}, Mouticello, IL, Sept. 2016

\bibitem{mlpeq}
S.~Chen, G.~Gibson, C.~Cown, and P.~Grant, ''Adaptive equalization of finite nonlinear channels using multilayer perceptrons,'' \emph{IEEE Trans. Signal Process.,} vol. 20, no. 2, pp. 107--119, Jun. 1990.




\bibitem{Imagenet}
A.~Krizhevsky, I.~Sutskever, and G.~E.~Hinton, ``Imagenet classification with deep convolutional neural networks,'' in \emph{Proc. Adv. Neural Inf. Process. Syst.}, 2012, pp. 1097-1105.
\bibitem{DL_NLP}
K.~Cho \emph{et al.} ``Learning phrase representations using {RNN} encoder-decoder for statistical machine translation.'' [Online]. Available: http://arxiv.org/abs/1406.1078, 2014
\bibitem{DL_Speech}
C.~Weng, D.~Yu, S.~Watanabe, and B.~H.~F.~Juang, ``Recurrent deep neural networks for robust speech recognition,'' in \emph{Proc. ICASSP}, May 2014, pp. 5532--5536.
\bibitem{winner}
P.~Kyosti, ``IST-4-027756 WINNER II D1.1.2 v.1.1: WINNER II Channel Models,'' 2007, [online] Available: http://www.ist-winner.org.
\bibitem{clipping}
X.~D.~Li and L.~J.~Cimini  Jr., ``Effects of clipping and filtering on the performance of OFDM,''  \emph{IEEE Comm. Lett.}, vol. 2, no. 5, pp. 131--133, May 1998
\end{thebibliography}
\end{document}